\documentclass{elsart}
\usepackage{graphicx}
\begin{document} 

\begin{frontmatter}
\title{Microwave Surface Impedance of
{Y$_1$Ba$_2$Cu$_3$O$_{7-\delta}$} crystals\\ 
Experiment and comparison to a $d$-wave model}
\author{T. Jacobs and S. Sridhar} 
\address{Physics Department, Northeastern University, Boston, MA 02115 }
\author{C. T. Rieck and K. Scharnberg}
\address{Institut fur Angewandte Physik, Universit\"at
Hamburg, Germany }
\author{T. Wolf and J. Halbritter}
\address{Forschungszentrum Karlsruhe Technik und Umwelt, Karlsruhe, Germany} 
%\date{} 
\date{\today } 
%\maketitle

\begin{abstract}
We present measurements of the microwave surface
resistance $R_s$ and the penetration depth $\lambda$ of
Y$_1$Ba$_2$Cu$_3$O$_{7-\delta}$ crystals.  At low $T$, $\lambda (T)$
obeys a polynomial behavior, while $R_{\mathrm{s}}$ displays a
characteristic non-monotonic $T-$dependence. A detailed comparison of
the experimental data is made to a model of $d$-wave superconductivity
which includes both elastic and inelastic scattering.  While the model
reproduces the general features of the experimental data, three
aspects of the parameters needed are worth noting. The elastic
scattering rate required to fit the data is much smaller than measured
from the normal state, the scattering phase shifts have to be close to
$\pi /2$ and a strong coupling value of the gap parameter $2\Delta
(0)/\mathrm{k}T_c\sim 6$ is needed. On the experimental side the
uncertainties regarding the material parameters $\lambda(0)$ and
$R_{\mathrm{s,\,res}}(0)$ further complicate a quantitative
comparison.  For one sample, $R_{\mathrm{s,\,res}}(0)$ agrees with the
intrinsic value which results from the $d$-wave model.
\end{abstract}
\end{frontmatter}

%\begin{keyword} high $T_{\mathrm{c}}$, surface resistance, penetration
%depth, {\it d}-wave theory \end{keyword}

%\subsection*{Introduction}

Microwave measurements of the surface impedance
$Z_{\mathrm{s}}=R_{\mathrm{s}}+\mathrm{i} X_{\mathrm{s}}$ of
superconductors are in principle capable of yielding a wealth of
precise information regarding the superconducting state, such as the
gap parameter, quasiparticle density and nature of scattering. In low
$T_{\mathrm{c}}$ superconductors the BCS theory provides a remarkably
accurate description of experimental data for $R_{\mathrm{s}}$ and
$X_{\mathrm{s}}$ over several orders of magnitude variation, including
detailed effects of impurity scattering \cite{Sridhar88a}.

Recently, experiments which directly explore the order parameter
symmetry suggest a $d_{x^2-y^2}$ order parameter \cite
{Wollman93,Tsuei94,Iguchi94} for high $T_{\mathrm{c}}$
superconductors, although some experiments which suggest a $s$-wave
order parameter also exist \cite {Chaudhari94,AGSun94}. It is
therefore useful to ask to what extent a $d$-wave model of
superconductivity can describe the measured surface impedance of the
cuprate superconductors, particularly
Y$_1$Ba$_2$Cu$_3$O$_{7-\delta}$. Some work has been already initiated
in this regard \cite{Borowski94}. Here in this paper, we present
detailed results on the microwave (10 GHz) surface impedance of
YBa$_2$Cu$_3$O$_{7-\delta}$ crystals. We also compare the complete
temperature dependence to numerical calculations based upon
semi-microscopic models of $d$-wave and $s$-wave superconductivity,
including elastic as well as inelastic scattering effects.

%\subsection*{Experimental Methods}

The measurements were carried out in a specially designed, high
sensitivity Nb cavity. The method of measuring the surface impedance
of superconductors at elevated temperatures using a ``hot finger''
cavity method was first introduced by one of the authors in
reference~\cite{Sridhar88}, and the principle has been used in a
variety of systems reported in the literature. In the present setup,
the Nb cavity is maintained either at $4.2\,{\mathrm{K}}$ or below
$2\,{\mathrm{K}}$. The typical background $Q_b$ of the cavity can be
as high as $10^8$. The surface resistance is measured from the
temperature dependent Q using $R_{\mathrm{s}}(T)=\Gamma
\,\,[Q^{-1}(T)-Q_b^{-1}(T)]$ and the penetration depth using $\Delta
\lambda (T)=\zeta \,[f(T)-f_b(T)]$. The geometric factors are
determined by the cavity mode, sample location and the sample
size. All measurements were done in the $TE_{\mathrm{011}}$ mode with
the sample at the midpoint of the cavity axis, where the microwave
magnetic fields have a maximum and the microwave electric fields are
zero.  The method enables measurement of small crystals and thin film
samples.

The crystals were grown in a ZrO$_2$/Y crucible using highly pure
Y$_2$O$_3$, BaCO$_3$ and CuO powders. Crystal growth took place while
slowly cooling the melt at a rate of $0.40 ^\circ \mathrm{C/h}$ in the
temperature range $970 ^\circ \mathrm{C}$ to $904 ^\circ \mathrm{C}$
and in an atmosphere of $100\,$mbar O$_2$. After the growth the
crystals were annealed in flowing oxygen in the temperature range $600
^\circ \mathrm{C}$ to $400 ^\circ \mathrm{C}$ during $600\,$h.

\begin{table} \caption{Some sample properties: $T_{\mathrm{c}}$ is
observed by us in small microwave fields. $\Delta T_{\mathrm{c}}$ is
defined as the temperature interval between $10\%$ and $90\%$ of
$R_{\mathrm{s}} (T_{\mathrm{c}})$. $\delta$ is the oxygen vacancy. }
\begin{tabular}{c|lrrr}
   Sample &$T_{\mathrm{c}}$&$\Delta T_{\mathrm{c}}$&$R_{\mathrm{s}}
        (4\,\mathrm{K})$& {$\delta$}\\ \hline
    A  &  $91.0\,\mathrm{K}$ &  $ 1.0\,\mathrm{K}$ & $ 5\,
 \mathrm{\mu \Omega}   $  &  $0.06$ \\
    B  &  $91.5\,\mathrm{K}$ &  $ 1.5\,\mathrm{K}$ & $150\,
 \mathrm{\mu \Omega}   $  &  $0.11$ \\
\end{tabular}
\label{tab:SampleProperties}
\end{table}

%\subsection*{Results}

%\paragraph{Normal State Properties:\ Scattering Times}

We start with an analysis of the normal state properties in the hope
to fix some of the normal state parameters which are required for a
calculation of the conductivity in the superconducting state. Assuming
local electrodynamics, {\it i.e.} skin depth limited, the normal state
surface resistance $R_{\mathrm{n}}$ is given by
$R_{\mathrm{n}}=\sqrt{\omega \mu _0\rho _{\mathrm{n}}/2}$ where $\rho
_{\mathrm{n}}=\sigma _0^{-1}=\mu _0\lambda(0)^2\,2\Gamma =(ne^2\tau
/m)^{-1}$. If the classical skin depth limit applies, then the
microwave resistivity $\rho _{\mathrm{n}}$ should be the same as the
dc resistivity, which is known to be linear. If $\rho
_{\mathrm{n}}=\rho _0+\gamma T$, then \begin{equation}
\label{rneqn}R_{\mathrm{n}}=\sqrt{\omega \mu _0(\rho _0+\gamma T)/2}
\end{equation}

\begin{figure} 
\includegraphics[angle=270,width=8cm]{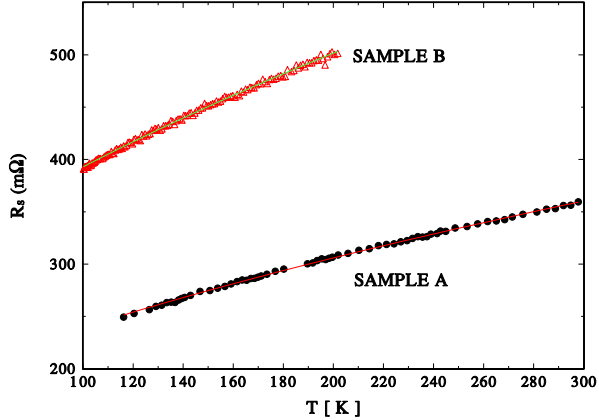}
\caption{Normal state surface resistance $R_{\mathrm{s}}(T)$
for the YBa$_2$Cu$_3$O$_{\mathrm{7-\delta}}$ crystals with fits to 
eq. (2). The fit parameters are given in table~2.}
\label{Fig1} 
\end{figure}

Figure~\ref{Fig1} shows a comparison of the experimental data between
$100\,\mathrm{K}$ and $200\,\mathrm{K}$ to the above equation. The
data clearly has a {\it sub-linear} $T$ dependence, and the fit to
eq.~(\ref{rneqn}) is {\em extremely good}.  From the fit the
scattering rate $\Gamma$ can be obtained as \begin{equation}
\label{scateqn1}\Gamma =\Gamma _{\mathrm{el}}+\Gamma
_{\mathrm{inel}}(T)=\Gamma _{\mathrm{el}}+\alpha T \end{equation}
where $\Gamma _{\mathrm{el}}=\rho _0/(2\mu _0\lambda(0)^2)$, and
$\alpha =\gamma /(2\mu _0\lambda
_0^2)$. Table~\ref{tab:ScatteringParameter} gives these parameters for
the studied samples under the assumption $\lambda(0) = 1400\,
\mathrm{\AA} $.

\begin{table} \caption{Experimental normal state scattering
parameters} \begin{tabular}{c|llll}
 {Sample} & $\rho_0$ & $\gamma$ & $\Gamma_{\mathrm{el}}$ & $\alpha$ \\
		                & $[\mathrm{\Omega m}]$   
    & $[\mathrm{\Omega m/K}]$    & $[\mathrm{sec}^{-1}]$  
       & $[\mathrm{sec}^{-1} \, \mathrm{K}^{-1}]$ \\ 
\hline
  A      &  $5.3 \, 10^{-7}$  &	$9.1 \, 10^{-9}$ &  $1.1 \, 10^{13}$	
&  $1.9 \, 10^{11}$  \\
  B      &  $1.4 \, 10^{-6}$  &  $2.4 \, 10^{-8}$ &  $2.9 \, 10^{13}$ 
   &  $5.0 \, 10^{11}$  \\
\end{tabular}
\label{tab:ScatteringParameter}
\end{table}

%\paragraph{Low Temperature Behavior of 
%$\lambda (T)$ and $R_{\mathrm{s}}(T)$}

\begin{figure} 
\includegraphics[width=8cm]{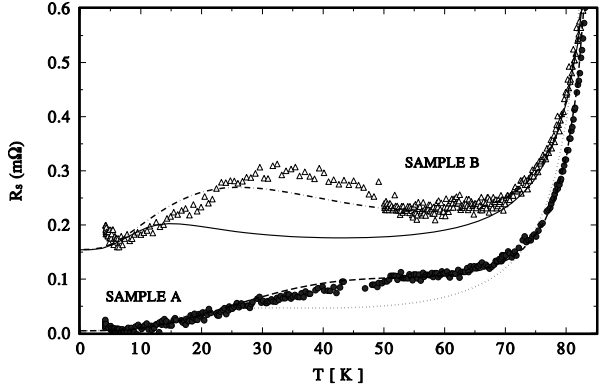}
\caption{
 Low temperature behavior of $R_{\mathrm{s}}$
 for the YBa$_2$Cu$_3$O$_{1-\delta}$ samples A ($\bullet$) and B
 ($\triangle$). Theoretical results are plotted for the parameters
 given in table~3.}
\label{Fig2} 
\end{figure}

Figure~\ref{Fig2} displays the low temperature behavior of
$R_{\mathrm{s}}$ for the two samples of
Y$_1$Ba$_2$Cu$_3$O$_{7-\delta}$. Sample B shows a characteristic peak
in $R_{\mathrm{s}}$ first reported by \cite{Bonn93}. This is not
present in sample A, whose behavior is instead closer to that of thin
films.

\begin{table} \caption{Theoretical parameter with $f_1(t) =
\mathrm{e}^{7(t-1)} -\mathrm{e}^{-7}$}
\begin{tabular}{c|ccccc}
    Sample                    
                & symbol 
                  & $\Gamma_{\mathrm{el}}\,[\mathrm{meV}]$
                    & $\Gamma_{\mathrm{inel}}\,[\mathrm{meV}]$
		      & $2\Delta_0 (0)/\mathrm{k}T_{\mathrm{c}}$
    \\ \hline
    A         
              & dashes
    	        & $0.25$
		  & $10.86\,f_1(t)$
                    & $6.4$
    \\
                      & dots
                        & $0.20$
                          & $10.91\,t^3$
                            & $6.0$    
                           
    \\
    B
            & dash dot   
	      & $0.05$
                & $24.15\,f_1(t)$
                  & $6.4$
    \\
                    & line
                      & $0.05$
                        & $24.15\,t^3$
                          & $7.0$                            

\end{tabular}
\label{tab:Fitparameter}
\end{table}

It is evident that the $\lambda$ vs. $T$ data displayed in figure~4
do not show the exponential dependence $\Delta \lambda (T)\propto
\exp (-T/T_{\mathrm{c}})$ expected for an isotropic $s$-wave
superconductor. Instead the data clearly have a polynomial
temperature dependence with a leading linear term
\cite{Hardy93,JMao95}.  This, together with the nonexponential
decrease of $R_{\mathrm{s}}$ at very low temperatures could be taken
as indication for $d$-wave pairing.

An important prediction for any superconducting state with nodes in
the gap is the presence of a finite residual conductivity $\sigma_{00}
$ due to elastic scattering \cite{PALee93,Hirschfeld94}.  Its value
depends on the particular pair state and, if determined
experimentally, could help to identify the type of pairing present.
For the $d$-wave state \begin{equation} \label{opeqn1}\Delta
(T,\pol{k}_F)=\Delta _0(T)\,\cos (2\phi ) \end{equation} commonly
studied for systems with cylindrical Fermi surfaces of circular cross
section one has \cite{PALee93,Hirschfeld94} $\sigma_{00}=ne^2/m\pi
\Delta_0$ at $T=0\,\mathrm{K}$.

This can be related to $\sigma _{\mathrm{n}}(T_{\mathrm{c}})$ by
$\sigma _{00}/\sigma _{\mathrm{n}}(T_{\mathrm{c}})=2\Gamma
(T_{\mathrm{c}})/\pi \Delta _0$. Using the values $2 \Gamma
(T_{\mathrm{c}})=33\,$meV, and $2\Delta
_0/{\mathrm{k}}T_{\mathrm{c}}=4.3$ we get $\sigma _{00}/ \sigma
_{\mathrm{n}}(T_{\mathrm{c}}) \sim 1$. Thus the residual conductivity
is comparable to the normal state conductivity at $T_{\mathrm{c}}$,
which is a surprisingly large value. Note that in BCS the conductivity
$\sigma _1\rightarrow 0$ as $T\rightarrow 0$.

The relation to $R_{\mathrm{s}}$ is obtained from the limiting result
when $\sigma _2\gg \sigma _1$, whence $R_{\mathrm{s}}(T\rightarrow
0)=\omega ^2\mu _0^2\lambda(0)^2\sigma _{00}/2$. This can also be
written as $R_{\mathrm{s}}/R_{\mathrm{n}}=0.5(\sigma _1/\sigma
_{\mathrm{n}})/(\sigma _2/\sigma _{\mathrm{n}})^{3/2}$. Since the
above relationship shows that $\sigma _1/\sigma _{\mathrm{n}}\sim 1$,
the reduction in $R_{\mathrm{s}}$ from its value at $T_{\mathrm{c}}$
is entirely due to the reduction in $\lambda $ from $\delta
_{\mathrm{n}} $ {\it i.e.} to the superfluid response. The intrinsic
residual surface resistance $R_{\mathrm{s,\,res}}(0)$ which results
from $\sigma_{00}$ is much smaller than the value measured for sample
B but is compatible with the low temperature data for sample A.

%\paragraph{Comparison to $d$-wave and $s$-wave theories}

Detailed numerical calculations based on the microscopic models for
$s$-wave and $d$-wave superconductivity
\cite{scharnberg78,RAKlemm88,scharnberg95} were carried out.

With suitable choices for the Eliashberg functions it is possible to
fit the measured surface resistances within the framework of isotropic
strong coupling theory over a wide range of temperature from
$T_{\mathrm{c}}$ down to $0.4\,T_{\mathrm{c}}$ \cite{scharnberg95},
highlighting the importance of inelastic scattering processes.  This
is consistent with earlier measurements over the same temperature
range \cite{Sridhar89}. However at lower temperatures the isotropic
gap should make its presence known in the form of an $\exp (-\Delta
/{\mathrm{k}}T)$ dependence, which is clearly not observed in either
$R_{\mathrm{s}}$ or $ \Delta \lambda$.

For this reason we focus in this paper on models of $d$-wave
superconductivity, including inelastic scattering.  Here we list the
main ingredients of the model - details of the model are described
elsewhere \cite{scharnberg95}.  A weak-coupling pairing interaction
was suitably chosen to give the $d_{x^2-y^2}$ order parameter $\Delta
(T,\pol{k}_F)$, see eq.~(\ref{opeqn1}).  The self-consistency equation
for the order parameter was solved to give the temperature dependence
of the amplitude $\Delta _0(T)$, and which is found to be very similar
to that for an isotropic order parameter except that $ 2\Delta
_0(0)/{\mathrm{k}}T_{\mathrm{c}}=4.29$ rather than $3.52$.

Elastic scattering is parametrized by a normal state scattering rate
$\Gamma _{\mathrm{el}}$ and a phase shift $\delta_{\mathrm{N}}$, which
can take any value between $0$ (Born approximation) and $\pi /2$
(Unitary limit).  Inelastic scattering in the superconducting state is
parametrized in the form \begin{equation}
\label{ineqn1}\Gamma_{\mathrm{inel}}(T) = \alpha T_{\mathrm{c}} f(t)
\end{equation} with some function $f$ of the reduced temperature
$t=T/T_{\mathrm{c}}$.

The inputs to the calculation of the surface impedance
$Z_{\mathrm{s}}= (2\mathrm{i}/\sigma_{\mathrm{s}})^{1/2}$ are then
$\Gamma _{\mathrm{el}}$, $\delta _{\mathrm{N}}$, $\alpha$, and $f(t)$.
In order to fit the steep drop of $R_{\mathrm{s}}$ below
$T_{\mathrm{c}}$ we have found it necessary to treat $2\Delta
_0(0)/{\mathrm{k}}T_{\mathrm{c}}$ as a variable parameter.

The surface impedance is not as sensitive to the choice of model
parameters as is the real part of the conductivity $\sigma_1$, which
is also of some intrinsic interest.  Unfortunately, the peak height of
the experimentally determined $\sigma_1(T,\omega )$ depends strongly
on the choice of the zero temperature penetration depth $\lambda(0)$
while the low temperature behavior of $\sigma_1$ can be changed
substantially by subtracting from the measured surface resistance
$R_{\mathrm{s}}$ a residual loss $R_{\mathrm{s}}^{\mathrm{o}}$
\cite{Bonn93}.  The choice of $R_{\mathrm{s}}^{\mathrm{o}}$ is limited
by the consideration that $\sigma_1(T=0,\omega )$ should neither be
negative nor should it exceed $\sigma_1(T_{\mathrm{c}},\omega )$ by a
wide margin.  With these restrictions we find
$R_{\mathrm{s}}^{\mathrm{o}} = 0.15\ \mathrm{m\Omega}$ for sample B
while for sample A $R_{\mathrm{s}}^{\mathrm{o}}$ is negligible.

\begin{figure} \includegraphics[width=8cm]{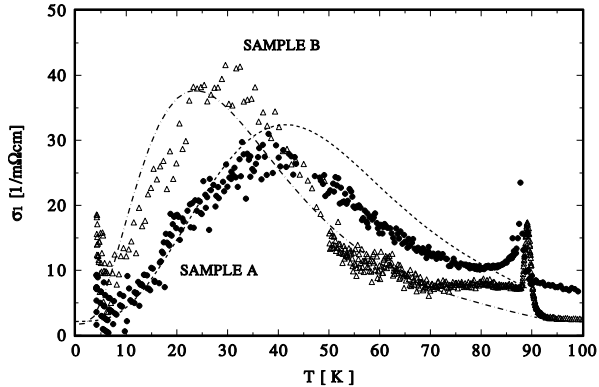} \caption{Real
part $\sigma_1$ of the measured conductivity and theoretical behavior
(lines). The parameters are given in table~3.}
\label{Fig3} \end{figure}

The peak heights in $\sigma_1$ are reduced when larger values for
$\lambda(0)$ are selected. We have chosen $\lambda(0)$ such that the
experimental $X_{\mathrm{s}}$ is linear over as large a temperature
range below $T_c$ as possible.  This gives $\lambda(0) = 1800\,
\mathrm{\AA}$ and $\sigma^{\mathrm{max}}_1(T,\omega ) = 30\, \lbrack
\mathrm{m\Omega cm}\rbrack^{-1}$ for sample A and $\lambda(0) = 2000\,
\mathrm{\AA}$ and $\sigma^{\mathrm{max}}_1(T,\omega ) = 42\, \lbrack
\mathrm{m\Omega cm}\rbrack^{-1}$ for sample B.

Experimental results for $\sigma_1$ are shown in figure~\ref{Fig3}.
In the presence of purely inelastic scattering a peak in $\sigma_1$
should occur near $\omega/2\Gamma (T) = 1$. This peak is thus expected
to shift to higher temperatures as the frequency is increased and to
lower temperatures when the overall magnitude $\alpha$
(eq.~(\ref{scateqn1}) and (\ref{ineqn1})) of the scattering rate is
increased. For $f(t)=f_1(t)=e^{7(t-1)} - e^{-7}$ which has been
suggested by Bonn {\it et al.} \cite{Bonn93} and which is close to the
result from the Nested Fermi Liquid model
\cite{Ruvalds90,scharnberg95}, the peak in $\sigma_1$ occurs near
$10\,\mathrm{K}$ with the peak height greatly exceeding the
experimental value. This observation, taken together with the normal
state data, shows that elastic scattering needs to be taken into
account.  When $\Gamma_{\mathrm{el}}$ given in table~2 is used as
input parameter, the peak in $\sigma_1$ is greatly reduced. In the
Born approximation $\sigma_1$ rises very steeply from its limiting
value $\sigma_{00}$ so that the peak is still located at too low a
temperature. In the unitary limit a peak is barely observable. In
order to reproduce the experimental results, $\Gamma_{\mathrm{el}}$
has to be chosen much smaller than the analysis of normal state data
would suggest.  Figure~\ref{Fig3} contains a reasonably close fit to
the $\sigma_1$ data.  The fit cannot be improved by varying the phase
shift $\delta_{\mathrm{N}}$.  Reducing $\delta_{\mathrm{N}}$ from
$0.5\pi$ shifts some of the weight of the peak shown in
figure~\ref{Fig3} to the temperature at which the peak occurs in the
Born approximation.  At around $\delta_{\mathrm{N}} = 0.35\pi$,
$\sigma_1(T)$ acquires a distinct double peak structure not compatible
with the data. We conclude that the phase shift must be close to the
unitary limit $0.45\pi \leq \delta_{\mathrm{N}} \leq 0.5\pi$.

The only way to improve the fit would be to choose different
temperature dependencies for $\Gamma_{\mathrm{inel}}$ with $f(t)$
decreasing faster then $f_1(t)=e^{7(t-1)} - e^{-7}$ for sample B and
more slowly for sample A. Even though the two samples differ in oxygen
contents (table~1) and by a factor of four in thickness, it does not
seem plausible that intrinsic scattering events in the two samples
should differ significantly in their temperature dependencies. A more
likely source for this discrepancy is a temperature dependence of the
``residual" surface resistance $R_{\mathrm{s,\,res}}(t)$.  Models have
been put forward to explain $R_{\mathrm{s,\,res}}(T)$ by weak links
and relate them to $\rho_0$ \cite{Halbritter93}.

The theoretical surface resistance for the same model parameters as
those used to calculate $\sigma_1$ is shown in figure~\ref{Fig2}.  In
the case of sample B the phenomenological residual resistance
$R_{\mathrm{s,\,res}}(0)=0.15 \,\mathrm{m\Omega}$ has been added. Note
the intrinsic residual surface resistance in the case of sample A.  To
fit the data near $T_{\mathrm{c}}$ we had to increase
$2\Delta_0(0)/\mathrm{k}T_{\mathrm{c}}$ to $6.4$. This is
substantially larger than the weak coupling value $4.29$, which has
important implications for conclusions regarding fluctuations.  The
overall fit to the data is very good, showing the same small
discrepancies already apparent in figure~\ref{Fig3}.

\begin{figure} \includegraphics[width=8cm]{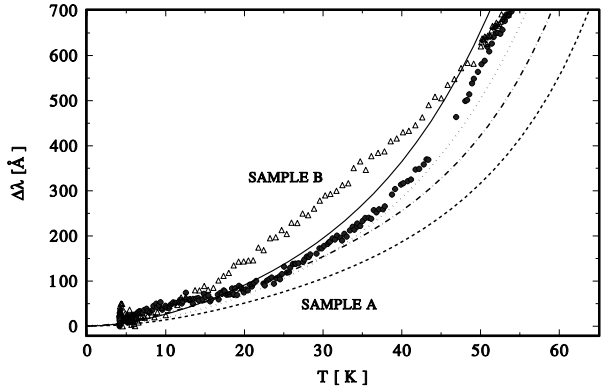} \caption{Low
temperature data and theory of $\lambda (T)$. Theoretical curves are
obtained with the parameters of table~3.  }  \label{Fig4} \end{figure}

The case for $d$-wave pairing would be strengthened considerably if we
could fit the shift in penetration depth equally well using exactly
the same model.  Experimental and theoretical results are compared in
figure~\ref{Fig4}. Clearly, the agreement is less than
satisfactory. {\it{Note that the calculated $\Delta \lambda$ is by no
means linear}}, although a polynomial fit in a limited temperature
range can certainly be found. The agreement could be much improved by
choosing different temperature dependences for
$\Gamma_{\mathrm{inel}}$. A good fit for sample
A is obtained with  $f_2(t) = t^3$, see figure~4.  Decreasing 
$2\Delta_0(0)/\mathrm{k}T_{\mathrm{c}}$
substantially would also improve agreement in the case of
$\Delta\lambda$ but would lead to serious discrepancies in the case of
$R_{\mathrm{s}}$.

\begin{figure} \includegraphics[width=8cm]{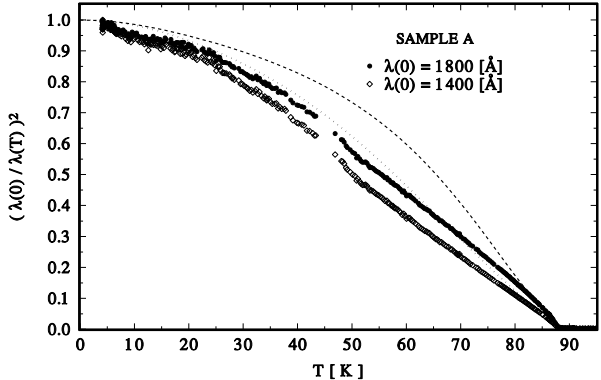}
\caption{Superfluid density of sample A. Dashed and doted lines are
obtained with the parameters of table~3.}  \label{Fig5} \end{figure}

In figure~\ref{Fig5} the penetration depth for sample A is plotted
over the whole temperature range in the form of the superfluid density
$(\lambda(0)/\lambda (T))^2$.  For such a plot, it is necessary to
assume a value of $\lambda(0)$ and this figure shows the variation of
$(\lambda(0)/\lambda (T))^2$ resulting from different choices of
$\lambda(0)$. By varying $\lambda(0)$ one can actually change the sign
of the curvature of $\lambda^2 (T))$ near $T_{\mathrm{c}}$.
 It is obvious that near $T_{\mathrm{c}}$, since a straight line fits
the data pretty well, the behavior is described quite well by a
$(1-T/T_{\mathrm{c}})^{1/2}$ dependence, which suggests a mean-field
behavior of the order parameter.

Our $d$-wave calculations give a positive curvature for $1/\lambda^2
(T)$ near $T_{\mathrm{c}}$, indicating a behavior which is slower than
mean-field, {\it i.e.} $\lambda (T)\rightarrow (1-t)^\nu $, where $\nu
>1/2$.  This is possibly an artifact of the calculation because the
total scattering rate (eq.~(\ref{scateqn1})) at $T_{\mathrm{c}}$ is
such that $T_{\mathrm{c}}$ should be substantially suppressed
according to weak coupling theory.  This would lead to noticeably
different $T_{\mathrm{c}}$'s for the two samples. Since the
$T_{\mathrm{c}}$'s are practically the same, we did not include
scattering in the selfconsistency equation.  Strong inelastic
scattering probably has to be treated within the framework of an
anisotropic strong coupling theory, which could also solve the problem
of the large value for $2\Delta_0(0)/\mathrm{k}T_{\mathrm{c}}$ we had
to assume.

In spite of the remaining discrepancies, some of which may be due to
contributions from the c-axis conductivity which has not been included
in the calculations, d-wave pairing seems to provide an adequate model
for understanding features seen in YBa$_2$Cu$_3$O$_{7-\delta}$
crystals. The main features of the data appear to be reproduced,
although a detailed microscopic justification of the needed parameters
is not yet available. We should remark that although we have
considered an explicit $d$-wave model, the essential feature is that
of nodes in the gap leading to low lying quasiparticle excitations at
all temperatures.

This work was supported by NSF-DMR-9223850. We thank M. Osofsky for
measurements of the oxygen content of the samples, as well as the
organizers of the International Symposium on HTSC in High Frequency
Fields held at Cologne, where part of this collaboration was
initiated.


\begin{thebibliography}{10}

\bibitem{Sridhar88a}
S.~Sridhar,
\newblock J. Appl. Phys. {\bf 63}, 159 (1988).

\bibitem{Wollman93}
D.~A. Wollman, D.~J. Van~Harlingen, W.~C. Lee, D.~M. Ginsberg, and A.~J.
  Leggett,
\newblock Phys. Rev. Lett. {\bf 71}, 2134 (1993).

\bibitem{Tsuei94}
C.~C. Tsuei, J.~R. Kirtley, C.~C. Chi, L.~S. Yu-Jahnes, A.~Gupta, T.~Shaw,
  J.~Z. Sun, and M.~Ketchen,
\newblock Phys. Rev. Lett. {\bf 73}, 593 (1994).

\bibitem{Iguchi94}
I.~Iguchi and Z.~Wen,
\newblock Phys. Rev. B {\bf 49}, 12388 (1994).

\bibitem{Chaudhari94}
P.~Chaudhari and S.-Y. Lin,
\newblock Phys. Rev. Lett. {\bf 72}, 1084 (1994).

\bibitem{AGSun94}
A.~G. Sun, D.~Gajewski, M.~Maple, and R.~C. Dynes,
\newblock Phys. Rev. Lett. {\bf 72}, 2267 (1994).

\bibitem{Borowski94}
L.~S. Borkowski and P.~J. Hirschfeld,
\newblock Phys. Rev. B {\bf 49}, 15404 (1994).

\bibitem{Sridhar88}
S.~Sridhar and W.~L. Kennedy,
\newblock Rev. Sci. Instrum. {\bf 59}, 531 (1988).

\bibitem{Bonn93}
D.~A. Bonn, K.~Zhang, R.~Liang, D.~Baar, and W.~N. Morgan, D. C.~andHardy,
\newblock J. Supcond. {\bf 6}, 219 (1993).

\bibitem{Hardy93}
W.~N. Hardy, D.~A. Bonn, D.~C. Morgan, R.~Liang, and K.~Zhang,
\newblock Phys. Rev. Lett. {\bf 70}, 3999 (1993).

\bibitem{JMao95}
J.~Mao, D.-H. Wu, J.~Peng, R.~L. Greene, and S.~M. Anlage,
\newblock Phys. Rev. B {\bf 51}, 3316 (1995).

\bibitem{PALee93}
P.~A. Lee,
\newblock Phys. Rev. Lett. {\bf 71}, 1887 (1993).

\bibitem{Hirschfeld94}
P.~J. Hirschfeld, W.~O. Putikka, and D.~J. Scalapino,
\newblock Phys. Rev. B {\bf 50}, 10250 (1994).

\bibitem{scharnberg78}
K.~Scharnberg,
\newblock J. Low Temp. Phys. {\bf 30}, 229 (1978).

\bibitem{RAKlemm88}
R.~A. Klemm, K.~Scharnberg, D.~Walker, and C.~T. Rieck,
\newblock Z. Phys. B {\bf 72}, 139 (1988).

\bibitem{scharnberg95}
C.~T. Rieck {\it et al},
\newblock Intrinsic surface impedance of weak and strong coupling
  superconductors: Temperature dependent scattering times and anisotropic
  energy gaps.,
\newblock to be published, 1995.

\bibitem{Sridhar89}
S.~Sridhar, D.-H. Wu, and W.~L. Kennedy,
\newblock Phys. Rev. Lett. {\bf 63}, 1873 (1989).

\bibitem{Ruvalds90}
A.~Virosztek and J.~Ruvalds,
\newblock Phys. Rev. B {\bf 42}, 4064 (1990).

\bibitem{Halbritter93}
J.~Halbritter,
\newblock Phys. Rev. B {\bf 48}, 9735 (1993).

\end{thebibliography}
\end{document}